\documentclass[preprint,aps,prd,showpacs,nofootinbib]{revtex4}
\usepackage{amssymb}%
\usepackage{graphicx}
\usepackage{subfigure}
\usepackage{amsmath}
\usepackage{bm}
\usepackage{hyperref}
\usepackage{relsize}
\RequirePackage{xspace}
\begin{document}


\title{\mbox{}\\[10pt]
Dijet Invariant Mass Distribution in Top Quark Hadronic Decay with
QCD Corrections}

\author{Hua-Sheng Shao$~^{(a)}$, Yu-Jie Zhang$~^{(b)}$ and Kuang-Ta
Chao$~^{(a ,\hspace{0.1cm}c)}$} \affiliation{ {\footnotesize
(a)~Department of Physics and State Key Laboratory of Nuclear
Physics and Technology, Peking University,
Beijing 100871, China}\\
{\footnotesize (b)~Key Laboratory of Micro-nano
 Measurement-Manipulation and Physics (Ministry of Education) and School of Physics, Beihang University,
 Beijing 100191, China}\\
{\footnotesize (c)~Center for High Energy Physics, Peking
University, Beijing 100871, China}\\
}





\begin{abstract}

The dijet invariant mass distributions from the hadronic decay of
unpolarized top quark ($t\to b W^+$ followed by $W^+\to u \bar{d}$)
are calculated, including the next-to-leading order QCD radiative
corrections. We treat the top decay in the complex mass scheme due
to the existence of the intermediate state W boson. Our analytical
expressions are also available in different dimensional
regularization schemes and $\gamma_5$ strategies. Finally, in order
to construct the jets, we use different jet algorithms to compare
their influences on our results. The obtained dijet mass
distributions from the top quark decay are useful to distinguish
these dijets from those produced via other sources and to clarify
the issue about the recent CDF Collaborations' $Wjj$ anomaly.

\end{abstract}

\maketitle


\section{Introduction}
Since the discovery of the top quark at the
Tevatron\cite{Abe:1995hr,Abachi:1995iq}, the top quark has played a
special role in searching for the electroweak symmetry breaking
mechanism and new physics beyond the standard model. This can be
attributed to the large mass of the top quark (about $173~{\rm GeV}$
), which is almost 40 times larger than the next heaviest quark. As
the Cabibbo-Kabayashi-Maskawa matrix element $V_{tb}$ approaches to
1, the top quark decays almost to a bottom quark and a W boson. Its
decay width
\cite{Harris:2002md,Jezabek:1988iv,Campbell:2004ch,Schmidt:1995mr}
is $\mathcal{O}(GeV)$, much larger than the typical QCD scale
$\Lambda_{QCD}\sim 300 MeV$, indicating that the top quark decay
takes place before hadronization. Therefore, nonperturbative effects
are not important in the properties of the top quark, and one can
perturbatively calculate its physical quantities precisely, such as
top quark's spin correlation. At the Large Hadron Collider (LHC) at
CERN, thousands of top quarks are expected to be produced per year
at 14 TeV. Hence a new era in top quark research has arrived.

On the other hand, very recently a dijet bump around 150 GeV in the
$Wjj$ channel has been observed by the Collider Detector at Fermilab
(CDF) at the Tevatron\cite{Aaltonen:2011mk}, and it has attracted a
lot of attention. There are some explanations within the standard
model for this anomaly\cite{He:2011ss,Sullivan:2011hu,Plehn:2011nx}.
Some studies may indicate that single top production may play an
important role in the CDF dijet excess. Moreover, the D0
Collaboration reported that their results were consistent with the
standard model's prediction in the same channel\cite{Abazov:2011af}.
Hence a careful investigation regarding the dijet in the single top
production and decay is helpful. Even without this CDF anomaly, it
is still useful to study the dijet distribution in the top quark
decay, as a part of investigations for the top quark properties.
Inspired by this, in the present study we will investigate the dijet
mass distribution in the top quark decay. This work also aims at
understanding the properties of top quarks.

There are a lot of works already about top quark
decays\cite{Denner:1990ns,Brandenburg:2002xr,Liu:1990py,Czarnecki:1990kv,
Li:1990qf,Jezabek:1988iv,Ghinculov:2000nx,Eilam:1991iz,Barroso:2000is,Oliveira:2001vw,
Czarnecki:1998qc,Chetyrkin:1999ju,Fischer:1998gsa,Fischer:2000kx,Fischer:2001gp,Penin:1998wj,Jezabek:1993wk,Do:2002ky}.
Generally, the QCD next-to-leading order (NLO) radiative corrections
to the top quark's width amount to about
$-8.54\%$\cite{Denner:1990ns,Liu:1990py,Czarnecki:1990kv,
Li:1990qf,Jezabek:1988iv,Ghinculov:2000nx}, while the corrections of
QCD two loops \cite{Czarnecki:1998qc,Chetyrkin:1999ju} and
electroweak one loop\cite{Denner:1990ns,Eilam:1991iz,Barroso:2000is}
are about $-2.05\%$ and $1.54\%$ respectively. The nonvanishing
$m_b$ effects
\cite{Fischer:1998gsa,Fischer:2000kx,Fischer:2001gp,Penin:1998wj}
and finite width corrections\cite{Jezabek:1993wk} reduce the Born
level width by about $0.27\%$ and $1.55\%$. All these show that the
QCD NLO corrections, among others, are important for top quarks.
Therefore, in the present study, we also put stress on the effects
of QCD NLO corrections to the dijet distributions in the top quark
decay.

The rest of the paper is organized as follows. Section
\uppercase\expandafter{\romannumeral 2} demonstrates the dimensional
regularization schemes and $\gamma_5$ schemes. Section
\uppercase\expandafter{\romannumeral 3} tackles our
scheme-independent analytical expressions. Jet algorithms are
recalled in Sec. \uppercase\expandafter{\romannumeral 4} and Sec.
\uppercase\expandafter{\romannumeral 5} discusses the results. The
final section contains the conclusion.

\section{Dimensional Regularization Schemes And $\gamma_5$ Schemes}
Dimensional regularization has many advantages in dealing with
ultraviolet, infrared and mass divergences encountered in high-order
calculations in a unified manner. However, there are still some
freedoms to handle these divergences in dimensional regularization.
In this section we recall four modern versions of frequently used
schemes and adopt the first three in the rest of this paper. The
four schemes include the conventional dimensional regularization
(CDR), the 't-Hooft-Veltman scheme (HV)\cite{'tHooft:1972fi}, the
four dimensional helicity scheme
(FDH)\cite{Bern:2002zk,Kunszt:1993sd,Bern:1991aq,Bern:1994zx,Bern:1994cg},
and the dimensional reduction scheme (DR)\cite{Siegel:1979wq}.

In CDR, only the $d=4-2\epsilon$ dimensional metric tensor is
introduced, i.e. $g^{\mu}_{\mu}=d$. The loop momentum and the spins
of vectors, regardless of whether they are "observed" or
"unobserved"\footnote{We call the hard and non-collinear external
particles observed states and internal, soft, or collinear external
particles unobserved states in this context.} are in $d$ dimensions,
whereas the spins of the spinor are in $d_s$ dimensions with
$d_s\geq d$. In this section, the observed states refer to the
external states appearing in the hard part of the process without
any subsequent hadronization. We treat $d_s$ of fermions as four
because it is distinct from $d$ and always appears as a global
factor in computations.

HV and FDH have many advantages in helicity amplitude calculations,
while FDH and DR are two supersymmmetric preserving
schemes\cite{Siegel:1979wq,Kunszt:1993sd,Bern:1994zx,Bern:1994cg}.
We describe the schemes in a unified way as explained below:

\begin{itemize}
\item To maintain the gauge invariance, all momentum integrals are
integrated in $d$ dimensions.

\item The dimensions of all observed particles (hard and noncollinear external
particles) are left in four dimensions.

\item The dimensions of all unobserved particles (internal
states and soft or collinear external states ) are treated in $d_s$
dimensions . Any explicit factors of dimension arising from these
state should be labeled as $d_s$ temporary; these must be kept
distinct from $d$ at the beginning.
\end{itemize}
We treat internal states with $d>4$ in HV and FDH, whereas $d<4$ in
DR, i.e., all variables in $d$ dimensions can be divided into a
four-dimensional part and $d-4$-dimensional part in CDR, HV and FDH,
while a four-dimensional quantity can be split into $d$ and
$4-d$-dimensional quantities in DR. The expressions are analytic
functions of $d$ and they are continued to any desired regions.
Setting $d_s=d$ denoted in the above items, we obtain the HV scheme,
while setting $d_s=4$ results in the FDH and DR schemes. As
mentioned above, the $d_s$ arising from dimensions of spinor space
is just a global factor. Therefore, we can set this part of $d_s$ to
be equal to 4. All of the above is summarized in Table
\ref{tab:dim}.

\begin{center}
\begin{table}[h]
\caption{\label{tab:dim} Summary of dimensions in different
regularization schemes.}
\end{table}
\begin{tabular}{c*{4}{c}}
\hline\hline \itshape ~Regularization schemes~ & \itshape ~~~CDR~~~
& \itshape ~~~HV~~~ & \itshape ~~~FDH~~~ & \itshape ~~~DR~~~
\\\hline
\itshape{\small Dimensions of momenta of observed particles} & $d$ &
$4$ & $4$ & $4$
\\\itshape{\small Dimensions of momenta of unobserved particles}& $d$ & $d$
& $d$ & $d$
\\ \itshape {\small Number of polarizations of observed massless vector bosons}& $d-2$  & $2$ & $2$ & $2$
\\ \itshape {\small Number of polarizations  of unobserved massless vector
bosons} & $d-2$ & $d-2$ & $2$ & $2$
\\ \itshape {\small Number of polarizations of observed massive vector
bosons} & $d-1$  & $3$ & $3$ & $3$
\\ \itshape {\small Number of polarizations of unobserved massive vector
bosons}& $d-1$ & $d-1$ & $3$ & $3$
\\ \itshape {\small Number of polarizations  of fermions}& $2$ & $2$
& $2$  & $2$
\\\hline\hline
\end{tabular}
\end{center}

Dimensional regularization has algebraic consistency problems with
respect to $\gamma_5$.
$\gamma_5=\frac{i}{4}\varepsilon_{\mu\nu\rho\sigma}
\gamma^{\mu}\gamma^{\nu}\gamma^{\rho}\gamma^{\sigma}$, which is well
defined in four dimensions. However, there are some problems with
this definition because antisymmetric tensor
$\varepsilon_{\mu\nu\rho\sigma}$ lives in four dimensions only. In
the naive definition of $\gamma_5$, some obviously inconsistent
equalities appear. If we keep all the four-dimensional rules and
cyclicity of the trace, the analytic continuation is
forbidden\cite{Kreimer:1989ke}. Therefore, one should at least
change one of the properties to obtain a consistent result. To the
best of our knowledge, there are two kinds of well-known $\gamma_5$
strategies that have been introduced; one is proposed by 't-Hooft
and Veltman and proved by Breitenlohner and Maison
\cite{Breitenlohner:1977hr,Breitenlohner:1975hg,Breitenlohner:1976te,'tHooft:1972fi}
(we call it the BMHV scheme), and the other one is introduced by
Korner, Kreimer and Schilcher
\cite{Kreimer:1993bh,Korner:1991sx,Kreimer:1989ke}(we call it the
KKS scheme).

As a compromise, in the BMHV scheme the anticommutation relationship
between $\gamma_5$ and $\gamma_{\mu}$ is violated, i.e.
$\{\gamma_5,\gamma_{\mu}\}\neq0$. In fact, every $d$-dimensional
quantity can be divided into a four-dimensional part and a
$-2\varepsilon$ part, which implies that in this scheme $d>4$.
$\gamma_5$ anticommutes with a four-dimensional $\gamma$-matrix,
while it commutes with a $-2\varepsilon$-dimensional
$\gamma$-matrix. This definition results in some ambiguousness of
chiral vector current treatment, e.g.
$\gamma^{\mu}\frac{1\pm\gamma_5}{2}\neq\frac{1\mp\gamma_5}{2}\gamma^{\mu}$
in tree-level Feynman rules. For the current work, we take the
symmetric version as presented in
\cite{Korner:1989is,Buras:1998raa}, i.e.
\begin{eqnarray}
\gamma^{\mu}\frac{1-\gamma_5}{2}&\rightarrow&\frac{1+\gamma_5}{2}\gamma^{\mu}\frac{1-\gamma_5}{2},\nonumber\\
\gamma^{\mu}\frac{1+\gamma_5}{2}&\rightarrow&-\frac{1+\gamma_5}{2}\gamma^{\mu}\frac{1-\gamma_5}{2}+\gamma^{\mu}.
\end{eqnarray}
The violation of anticommutation is also a violation of the Ward
identity in axial-vector currents. To prevent such a violation,
additional renormalization is needed\cite{Harris:2002md} (Readers
who are interested in dimensional renormalization issues can also
refer to
Refs.\cite{Martin:1999cc,Schubert:1988ke,Pernici:1999nw,Pernici:1999ga,Pernici:2000an,Ferrari:1994ct}).
This will be used in the next section. Although it is the first
rigorously proven consistent scheme, the process of isolating
four-dimensional and $-2\varepsilon$ parts in the Lorentz space
often suffers from complex practical calculations.

On all accounts, the strategy of covariance violation in $\gamma_5$
has some disadvantages in complicated situations. On the other hand,
the KKS scheme keeps the covariant anticommutations but forbids the
cyclicity in the trace. In $\gamma$-matrix algebra (Clifford
algebra), there is a unique generator, which anticommutes with all
other generators in infinite dimensions. This generator can be
defined as the $\gamma_5$. To avoid the cyclicity in the trace, the
"reading point" must be chosen first, and all $\gamma_5$ are moved
to this point before a trace is taken. This compromise recovers a
correct anomaly as well.

Finally, we also introduce the renormalization constants and the
splitting functions in the CDR, HV, and FDH dimensional
regularization schemes used in this paper. In order to avoid
calculating external self energy diagrams, we choose the on-shell
scheme for external legs. These constants are
\begin{eqnarray}
\delta Z^{OS}_{t}&=&-\frac{\alpha_s
C_F}{4\pi}\left(\frac{1}{\epsilon_{UV}}+\frac{2}{\epsilon_{IR}}-3\gamma_{E}+3\ln\left(\frac{4\pi\mu^2}{m_t^2}\right)
+4+1_{FDH}\right),\nonumber\\
\delta Z^{OS}_{q}&=&-\frac{\alpha_s
C_F}{4\pi}\left(\frac{1}{\epsilon_{UV}}-\frac{1}{\epsilon_{IR}}\right),
\end{eqnarray}
where $\delta Z^{OS}_{t}$, $\delta Z^{OS}_{q}$ are on-shell(OS) wave
function renormalization constants for top quark and light quarks,
respectively, $\gamma_{E}$ is the Euler constant, and $1_{FDH}$ is
only nonvanishing in FDH scheme. The unpolarized Altarelli-Parisi
splitting functions
\cite{Altarelli:1977zs,Catani:1996pk}\footnote{Our equations are the
same as those in ref.\cite{Catani:1996pk}. The discrepancies in the
$\mathcal{O}(\epsilon)$ parts and
refs.\cite{Kunszt:1993sd,Giele:1991vf,Giele:1993dj} were carefully
discussed in ref.\cite{Catani:1996pk}.} to $\mathcal{O}(\epsilon)$
in HV and CDR schemes are all listed in the following:
\begin{eqnarray}
P_{qq}(z)&=&C_F\frac{1+z^2}{1-z}-\epsilon~C_F(1-z),\nonumber\\
P_{gq}(z)&=&C_F\frac{1+(1-z)^2}{z}-\epsilon~C_F~z,\nonumber\\
P_{gg}(z)&=&2N_c\left(\frac{z}{1-z}+\frac{1-z}{z}+z(1-z)\right),\nonumber\\
P_{qg}(z)&=&\frac{z^2+(1-z)^2}{2}-\epsilon~z(1-z),
\end{eqnarray}
while in FDH and DR these terms should be
\begin{eqnarray}
P_{qq}(z)&=&C_F\frac{1+z^2}{1-z},\nonumber\\
P_{gq}(z)&=&C_F\frac{1+(1-z)^2}{z},\nonumber\\
P_{gg}(z)&=&2N_c\left(\frac{z}{1-z}+\frac{1-z}{z}+z(1-z)\right)+\epsilon~2N_c~z(1-z),\nonumber\\
P_{qg}(z)&=&\frac{z^2+(1-z)^2}{2}-\epsilon~z(1-z).
\end{eqnarray}
\section{Scheme Independence And Analytical Expressions}

As emphasized in Sec.\uppercase\expandafter{\romannumeral 2}, there
are some degrees of freedom to regularize possible divergences.
Because of unitarity in QCD cross sections\cite{Catani:1996pk}, we
should expect the scheme independence of the well-defined physical
results. In this section, analytical results are provided for top
quark decay and subsequent hadronic decay, thus affirming the
simplicity of these processes. Moreover, we also demonstrate that
the off-shell effect in the top quark hadronic decay is small, and
narrow-width-approximation is good enough at the decay width level.
\subsection{Corrections To $t\rightarrow b W^+$}
\begin{center}
\begin{figure}
\hspace{0cm}\includegraphics[width=19cm]{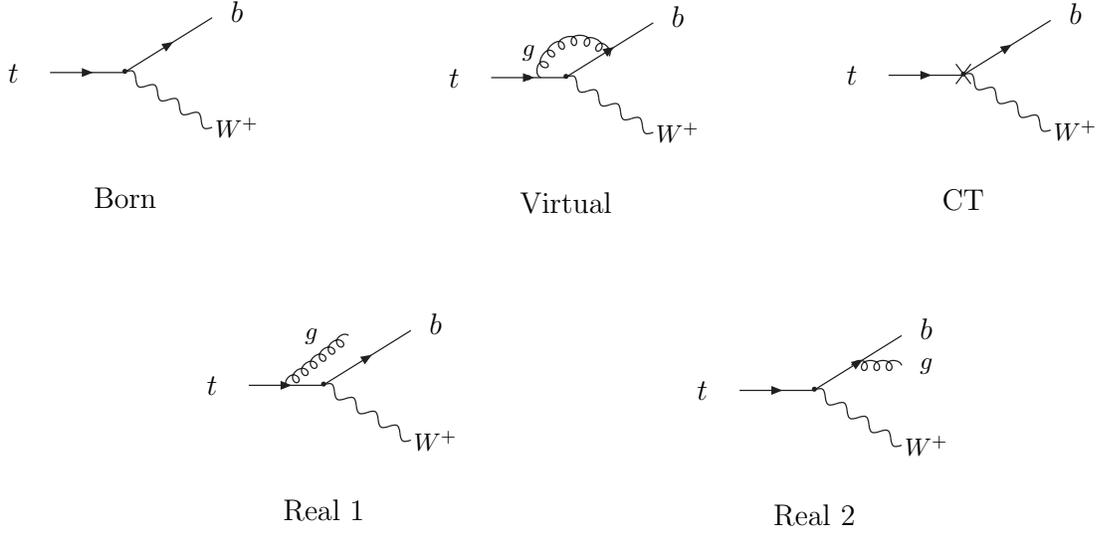}
\caption{\label{fig:graph1} Feynman diagrams in $t \rightarrow b
W^+$.}
\end{figure}
\end{center}

We first reproduce the well-known QCD corrections to the top quark
decay\cite{Harris:2002md,Jezabek:1988iv,Campbell:2004ch,Schmidt:1995mr}(Feynman
diagrams generated by FEYNARTS \cite{Hahn:2000kx} are shown in
Fig.\ref{fig:graph1}). Because of the Cabibbo-Kabayashi-Maskawa
(CKM) matrix elements $1\approx|V_{tb}|\gg|V_{ts}|,|V_{td}|$, the
branching ratio of $t\rightarrow b W^+$ is almost $100\%$. For
simplification, we set the CKM matrix to be diagonal and the mass of
the b-quark equal to zero. As presented in previous works, the
effect of nonvanishing mass of the bottom quark is negligible.
Following the notations of Ref.\cite{Campbell:2004ch}, the matrix
element of the tree-level process $t(p_t)\rightarrow b(p_b)
W^+(p_W)$  with averaging over the top quark's spin and color is
given by
\begin{eqnarray}
|\overline{\mathcal{M}}_0|^2&=&\frac{e^2m_t^4}{4s_w^2m_W^2}\left(1-r^2\right)\left(1+2r^2\right),
\end{eqnarray}
where $r=\frac{m_W}{m_t}$ and $s_w$ is the sine of Weinberg angle.
We can get the leading-order width easily
\begin{eqnarray}
\Gamma_0&=&\frac{\alpha
m_t^3}{16s_w^2m_w^2}\left(1-r^2\right)^2\left(1+2r^2\right),
\end{eqnarray}
where we have used the electromagnetic fine-structure coupling
constant $\alpha=\frac{e^2}{4\pi}$.

To check the regularization scheme independence of these results, we
first derive the averaged squared matrix element in the FDH
regularization scheme within the naive or KKS $\gamma_5$ scheme. The
virtual terms and counter-terms for renormalization are given by
\begin{eqnarray}
\left(|\overline{\mathcal{M}}_v|^2+|\overline{\mathcal{M}}_{ct}|^2\right)^{KKS}_{FDH}&=&|\overline{\mathcal{M}}_0|^2
\frac{\alpha_s~C_F}{2\pi\Gamma(1-\epsilon)}\left(\frac{4\pi\mu^2}{m_t^2}\right)^{\epsilon}\nonumber\\&&
\left[-\frac{1}{\epsilon^2}-\frac{\frac{5}{2}-2\ln(1-r^2)}{\epsilon}-\frac{11}{2}-\frac{\pi^2}{6}+3\ln(1-r^2)\right.
\nonumber\\&&-\frac{\ln(1-r^2)}{r^2}+\frac{1-r^2}{r^2(1+2r^2)}\ln(1-r^2)\nonumber\\&&\left.
-2\ln^2(1-r^2)-2~\text{Li}_2(r^2)\right].
\end{eqnarray}
In order to see the scheme-dependent terms, we subtract the
expressions in other schemes by the expressions in FDH with KKS
$\gamma_5$ treatment and use
$\delta|\overline{\mathcal{M}}_{v/ct/real}|^2\doteq|\overline{\mathcal{M}}_{v/ct/real}|^2
-\left(|\overline{\mathcal{M}}_{v/ct/real}|^2\right)^{KKS}_{FDH}$.
These scheme-dependent terms are
\begin{eqnarray}
\left(\delta|\overline{\mathcal{M}}_v|^2+\delta|\overline{\mathcal{M}}_{ct}|^2\right)^{KKS}_{HV}&=&-|\overline{\mathcal{M}}_0|^2
\frac{\alpha_s~C_F}{4\pi},\nonumber\\
\left(\delta|\overline{\mathcal{M}}_v|^2+\delta|\overline{\mathcal{M}}_{ct}|^2\right)^{KKS}_{CDR}&=&|\overline{\mathcal{M}}_0|^2
\frac{\alpha_s~C_F}{4\pi\Gamma(1-\epsilon)}\left(\frac{4\pi\mu^2}{m_t^2}\right)^{\epsilon}\frac{1}{1+2r^2}\nonumber\\&&
\left[\frac{4r^2}{\epsilon}+8r^2-1-8r^2\ln(1-r^2)\right],\nonumber\\
\left(\delta|\overline{\mathcal{M}}_v|^2+\delta|\overline{\mathcal{M}}_{ct}|^2\right)^{BMHV}_{FDH}&=&|\overline{\mathcal{M}}_0|^2
\frac{\alpha_s~C_F}{2\pi},\nonumber\\
\left(\delta|\overline{\mathcal{M}}_v|^2+\delta|\overline{\mathcal{M}}_{ct}|^2\right)^{BMHV}_{HV}&=&|\overline{\mathcal{M}}_0|^2
\frac{3~\alpha_s~C_F}{4\pi},\nonumber\\
\left(\delta|\overline{\mathcal{M}}_v|^2+\delta|\overline{\mathcal{M}}_{ct}|^2\right)^{BMHV}_{CDR}&=&|\overline{\mathcal{M}}_0|^2
\frac{\alpha_s~C_F}{4\pi\Gamma(1-\epsilon)}\left(\frac{4\pi\mu^2}{m_t^2}\right)^{\epsilon}\frac{1}{1+2r^2}\nonumber\\&&
\left[\frac{4r^2}{\epsilon}+3+16r^2-8r^2\ln(1-r^2)\right].
\end{eqnarray}

These scheme-dependent terms should be canceled  exactly with real
corrections originated from soft and collinear regions. In process
$t(p_t)\rightarrow b(p_b) W^+(p_W) g(p_g)$, the real correction
expressions in different schemes after integrating over the momentum
of the radiative gluon are given by
\begin{eqnarray}
\left(|\overline{\mathcal{M}}_{real}|^2\right)^{KKS}_{FDH}&=&|\overline{\mathcal{M}}_0|^2
\frac{\alpha_s~C_F}{2\pi\Gamma(1-\epsilon)}\left(\frac{4\pi\mu^2}{m_t^2}\right)^{\epsilon}\nonumber\\&&
\left[\frac{1}{\epsilon^2}+\frac{\frac{5}{2}-2\ln(1-r^2)}{\epsilon}-\frac{5\pi^2}{6}
-\frac{2(7r^4-5r^2-4)}{\left(1+2r^2\right)\left(1-r^2\right))}\right.\nonumber\\&&
-5\ln(1-r^2)+2\ln^2(1-r^2)\nonumber\\&&
-\left.\frac{2r^2(1+r^2)(1-2r^2)}{(1+r^2)^2(1+2r^2)}\ln(r^2)+2~\text{Li}_2(1-r^2)\right],\nonumber\\
\left(\delta|\overline{\mathcal{M}}_{real}|^2\right)^{BMHV}_{FDH}&=&0,\nonumber\\
\left(\delta|\overline{\mathcal{M}}_{real}|^2\right)^{KKS/BMHV}_{HV}&=&|\overline{\mathcal{M}}_0|^2\frac{\alpha_s~C_F}{4\pi}=
-\left(\delta|\overline{\mathcal{M}}_v|^2+\delta|\overline{\mathcal{M}}_{ct}|^2\right)^{KKS}_{HV},\nonumber\\
\left(\delta|\overline{\mathcal{M}}_{real}|^2\right)^{KKS/BMHV}_{CDR}&=&|\overline{\mathcal{M}}_0|^2
\frac{\alpha_s~C_F}{4\pi\Gamma(1-\epsilon)}\left(\frac{4\pi\mu^2}{m_t^2}\right)^{\epsilon}\frac{1}{1+2r^2}\nonumber\\&&
\left[-\frac{4r^2}{\epsilon}-8r^2+1+8r^2\ln(1-r^2)\right]\nonumber\\&=&
-\left(\delta|\overline{\mathcal{M}}_v|^2+\delta|\overline{\mathcal{M}}_{ct}|^2\right)^{KKS}_{CDR}.
\end{eqnarray}

Combining all the results above, we find that the results in the
three-dimensional regularization schemes in the KKS $\gamma_5$
strategy are the same; however these are not consistent with the
BMHV $\gamma_5$ scheme at present. In the BMHV $\gamma_5$ scheme,
the violation of anticommutation also violates the Ward identities,
which is also pointed out in Ref.\cite{Harris:2002md}. Furthermore,
to maintain the Ward identities, finite renormalization is made for
axial-vector currents,
\begin{eqnarray}
\left(\Gamma^{ren}_{\mu5}\right)_{FDH}&=&\left(1-\frac{\alpha_s~
C_F}{2\pi}\right)\Gamma^{bare}_{\mu5},\nonumber\\
\left(\Gamma^{ren}_{\mu5}\right)_{HV/CDR}&=&\left(1-\frac{\alpha_s~
C_F}{\pi}\right)\Gamma^{bare}_{\mu5},
\end{eqnarray}
where $\Gamma_{\mu5}$ represents the axial-vector current.

Thus far, we get the unique result\footnote{In general, we should
include finite renormalization terms of coupling constants in FDH
related to conventional $\overline{\rm MS}$ scheme
\cite{Kunszt:1993sd} to obtain the unique physical result. However,
all of our processes under consideration are only
$\mathcal{O}(\alpha_s)$ at the QCD one-loop level. This finite
renormalization is absent in our calculations.}
\begin{eqnarray}
\Gamma&=&\Gamma_0\left\{1-\frac{\alpha_s~C_F}{2\pi}\left[\frac{2\pi^2}{3}-\frac{3}{2}
-\frac{4}{3(1-r^2)}+\frac{1}{3(1+2r^2)}-2\ln(\frac{r^2}{1-r^2})\right.\right.\nonumber\\&&
+2\ln(r^2)\ln(1-r^2)+\frac{22-34r^2}{9(1-r^2)^2}\ln(r^2)
+\frac{3\ln(1-r^2)}{1+2r^2}\nonumber\\&&
-\left.\left.\frac{4\ln(r^2)}{9(1+2r^2)}+4~\text{Li}_2(r^2)\right]\right\}.
\end{eqnarray}
If we set $r\approx0.46$, we get the well-known K factor
$(1-0.8\alpha_s)$.
\subsection{Corrections To $W^+\rightarrow u \bar{d}$}
\begin{center}
\begin{figure}
\hspace{0cm}\includegraphics[width=19cm]{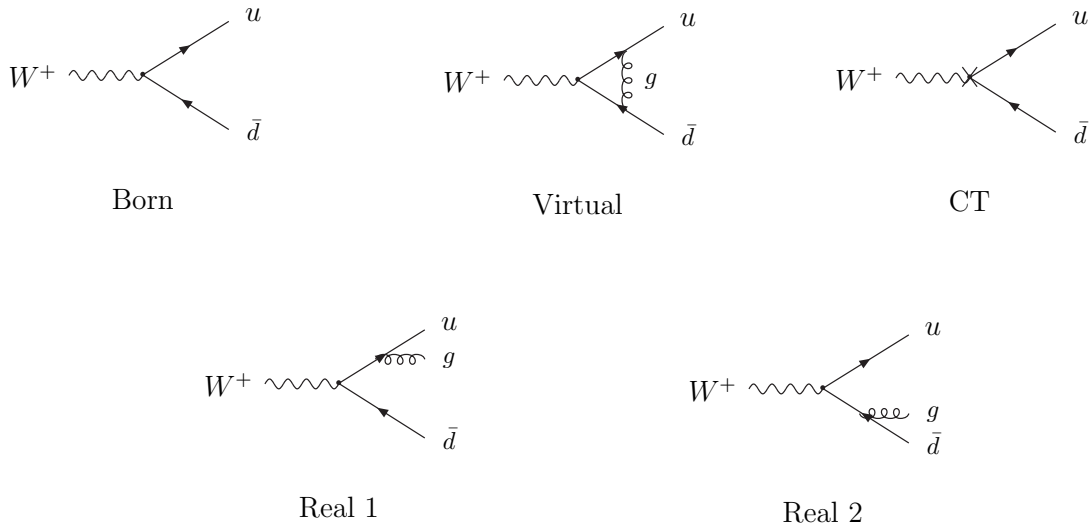}
\caption{\label{fig:graph2} The Feynman diagrams in $W^+ \rightarrow
u \bar{d}$.}
\end{figure}
\end{center}

With the same procedure described in the previous subsection, we
obtain the analytical results for the subsequent decay of W
boson\cite{Altarelli:1979ub,Denner:1990tx}. We labeled the momenta
of the W boson, up (charm) quark, and down (strange) quark as $p_W,
p_u, p_d$ respectively. The diagonalization of CKM matrix and
vanishing mass of light quarks guarantee a factor of 2 to the W
boson's hadronic decay channel via the process $W^+\rightarrow u
\bar{d}$. The diagrams of QCD correction to this process are all
shown in Fig.\ref{fig:graph2}.

The lowest-order squared matrix element and decay width are
\begin{eqnarray}
|\overline{\mathcal{M}}_0|^2&=&\frac{e^2m_W^2}{s_w^2},\nonumber\\
\Gamma_0&=&\frac{\alpha~m_W}{4~s_w^2}.
\end{eqnarray}

The contributions of virtual terms and counter-terms are
\begin{eqnarray}
\left(|\overline{\mathcal{M}}_v|^2+|\overline{\mathcal{M}}_{ct}|^2\right)^{KKS}_{FDH}&=&|\overline{\mathcal{M}}_0|^2
\frac{\alpha_s~C_F}{2\pi\Gamma(1-\epsilon)}\left(\frac{4\pi\mu^2}{m_W^2}\right)^{\epsilon}
\left(-\frac{2}{\epsilon^2}-\frac{3}{\epsilon}+\pi^2-7\right),\nonumber\\
\left(\delta|\overline{\mathcal{M}}_v|^2+\delta|\overline{\mathcal{M}}_{ct}|^2\right)^{KKS}_{HV}&=&-|\overline{\mathcal{M}}_0|^2
\frac{\alpha_s~C_F}{2~\pi},\nonumber\\
\left(\delta|\overline{\mathcal{M}}_v|^2+\delta|\overline{\mathcal{M}}_{ct}|^2\right)^{KKS}_{CDR}&=&|\overline{\mathcal{M}}_0|^2
\frac{\alpha_s~C_F}{\pi\Gamma(1-\epsilon)}\left(\frac{4\pi\mu^2}{m_W^2}\right)^{\epsilon}
\left(\frac{1}{\epsilon}+1\right),\nonumber\\
\left(\delta|\overline{\mathcal{M}}_v|^2+\delta|\overline{\mathcal{M}}_{ct}|^2\right)^{BMHV}_{FDH/HV}&=&-|\overline{\mathcal{M}}_0|^2
\frac{\alpha_s~C_F}{2~\pi},\nonumber\\
\left(\delta|\overline{\mathcal{M}}_v|^2+\delta|\overline{\mathcal{M}}_{ct}|^2\right)^{BMHV}_{CDR}&=&|\overline{\mathcal{M}}_0|^2
\frac{\alpha_s~C_F}{\pi\Gamma(1-\epsilon)}\left(\frac{4\pi\mu^2}{m_W^2}\right)^{\epsilon}
\left(\frac{1}{\epsilon}+2\right).
\end{eqnarray}

For real corrections after the phase space integration over
radiative gluon momentum, we arrive at
\begin{eqnarray}
\left(|\overline{\mathcal{M}}_{real}|^2\right)^{KKS}_{FDH}&=&|\overline{\mathcal{M}}_0|^2
\frac{\alpha_s~C_F}{2\pi\Gamma(1-\epsilon)}\left(\frac{4\pi\mu^2}{m_W^2}\right)^{\epsilon}
\left(\frac{2}{\epsilon^2}+\frac{3}{\epsilon}+\frac{17}{2}-\pi^2\right),\nonumber\\
\left(\delta|\overline{\mathcal{M}}_{real}|^2\right)^{BMHV}_{FDH}&=&0,\nonumber\\
\left(\delta|\overline{\mathcal{M}}_{real}|^2\right)^{KKS/BMHV}_{HV}&=&|\overline{\mathcal{M}}_0|^2
\frac{\alpha_s~C_F}{2\pi}
=-\left(\delta|\overline{\mathcal{M}}_v|^2+\delta|\overline{\mathcal{M}}_{ct}|^2\right)^{KKS}_{HV},\nonumber\\
\left(\delta|\overline{\mathcal{M}}_{real}|^2\right)^{KKS/BMHV}_{CDR}&=&|\overline{\mathcal{M}}_0|^2
\frac{\alpha_s~C_F}{\pi\Gamma(1-\epsilon)}\left(\frac{4\pi\mu^2}{m_W^2}\right)^{\epsilon}
\left(-\frac{1}{\epsilon}-1\right)\nonumber\\
&=&-\left(\delta|\overline{\mathcal{M}}_v|^2+\delta|\overline{\mathcal{M}}_{ct}|^2\right)^{KKS}_{CDR}.
\end{eqnarray}

After including the renormalization of the axial-vector current in
the BMHV $\gamma_5$ scheme, we obtain the scheme-independent answer
for the decay width of process $W^+\rightarrow u \bar{d}$,
\begin{eqnarray}
\Gamma=\Gamma_0\left(1+\frac{\alpha_s}{\pi}\right).
\end{eqnarray}

\subsection{Corrections To $t\rightarrow b u \bar{d}$}
\begin{center}
\begin{figure}
\hspace{0cm}\includegraphics[width=19cm]{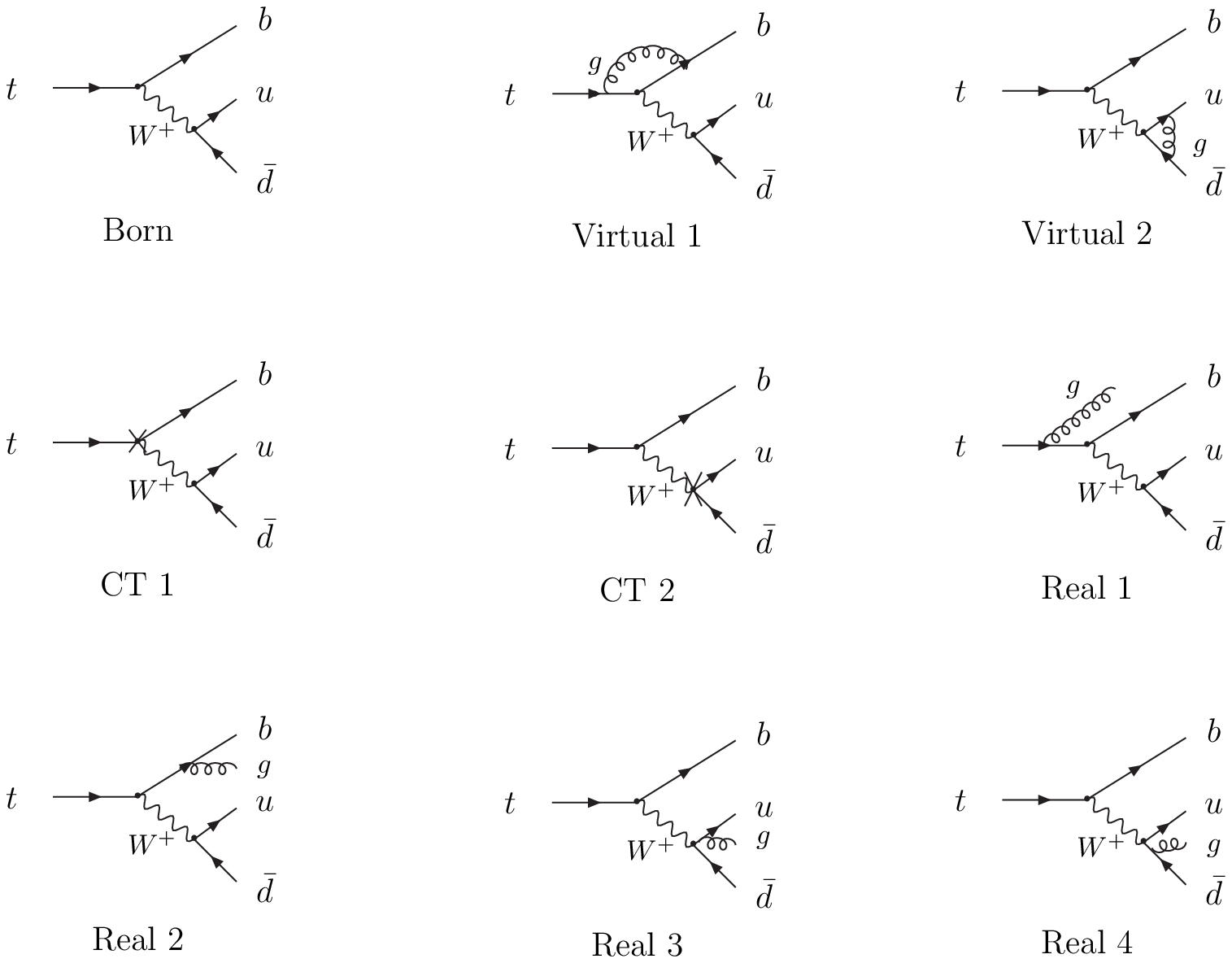}
\caption{\label{fig:graph3} Feynman diagrams in $t \rightarrow b u
\bar{d}$.}
\end{figure}
\end{center}

In this subsection, we present the analytical expressions of the top
quark hadronic decay. The corresponding graphs are shown in
Fig.\ref{fig:graph3}. As the mass of the top quark is 30 times
larger than that of the bottom quark, we set the masses of all final
states to be zero. The effect of the nonzero mass of the bottom
quark is negligible in our results. Because of the
intermediate-state $W$ boson in $t(p_t)\rightarrow b W^+\rightarrow
b(p_b) u(p_u) \bar{d}(p_d)$, we treat this process in the complex
mass scheme\cite{Denner:2005es,Denner:2005fg}. The Born amplitude
squared with averaging over the initial-state spin and color is
given by
\begin{eqnarray}
|\overline{\mathcal{M}}_{0}|^2=\frac{3e^4}{2|s_w|^4}\frac{(1-y)y}{(1-y-z-r^2)^2+(r^2w)^2},
\end{eqnarray}
where we have defined $(p_t-p_d)^2=m_t^2y,(p_t-p_u)^2=m_t^2z,
r=\frac{m_W}{m_t}, w=\frac{\Gamma_W}{m_W}$. Here, $y, z, r, w$ are
all dimensionless variables. We keep the width of the W boson
nonvanishing. The Born level decay width of this channel is
\begin{eqnarray}
\Gamma_0&=&\frac{\alpha^2m_t}{64\pi|s_w|^4}\left[4r^2-2+2r^2\left(w^2r^2-3r^2+3\right)\ln(r^2)\right.\nonumber\\&&
+r^2\left(w^2r^2-3r^2+3\right)\ln(\frac{1+w^2}{\left(1-r^2\right)^2+\left(w
r^2\right)^2})\nonumber\\&&\frac{6w^2r^6-2r^6-3w^2r^4+3r^4-1}{wr^2}\nonumber\\&&\left.\left(\tan^{-1}(w)+\tan^{-1}(\frac{w
r^2}{1-r^2})-\pi\right)\right].
\end{eqnarray}
By expanding it in terms of $w$, to $\mathcal{O}{\left(w^0\right)}$
the above result can be expressed as
\begin{eqnarray}
\Gamma_0&=&\frac{\alpha^2m_t}{64|s_w|^4r^2}\left[\left(1-r^2\right)\left(1+2r^2\right)w^{-1}+
\right.\nonumber\\&&
\left.\frac{\left(6r^4(1-r^2)\ln(\frac{r^2}{1-r^2})+6r^4-3r^2-1\right)}{\pi}w^0+\mathcal{O}(w^1)\right].
\end{eqnarray}
To leading order in $w$ the result is consistent with the
narrow-width-approximation (NWA), $\Gamma_{t\rightarrow b u
\bar{d}}=\Gamma_{t \rightarrow b W^+}\times\frac{\Gamma_{W^+
\rightarrow u \bar{d}}}{\Gamma_W}$, with the Born width formulas of
the top quark and W boson exhibited in two previous subsections. The
second term is an off-shell correction, which is about $-0.6 w$
relative to the first term with $r\approx0.46$.

In naive or KKS $\gamma_5$ strategy within FDH regularization
scheme, at QCD one loop level\footnote{Because of color flow,the W
boson propagator is not involved in loops. The scalar one loop
integrals with real masses encountered in this process were already
illustrated in Ref.\cite{Ellis:2007qk}. However, some analytical
continuations should be made in calculating scalar one-loop
integrals with complex arguments contrast to the ones with real
arguments.} the squared matrix elements after renormalization with
the initial-state averaged is given by
\begin{eqnarray}
\left(|\overline{\mathcal{M}}_v|^2+|\overline{\mathcal{M}}_{ct}|^2\right)^{KKS}_{FDH}&=&|\overline{\mathcal{M}}_{0}|^2
\frac{\alpha_s~C_F}{4\pi\Gamma(1-\epsilon)}\left(\frac{4\pi\mu^2}{m_t^2}\right)^{\epsilon}
\nonumber\\&&\left[-\frac{6}{\epsilon^2}+\frac{4\ln((y+z)(1-y-z))-11}{\epsilon}+6\ln((y+z)(1-y-z))\right.\nonumber\\&&
-2\ln^2(1-y-z)+4\ln(1-y-z)\ln(y+z)-4\ln^2(y+z)\nonumber\\&&\left.-\frac{2\ln(y+z)}{1-y}+4~\text{Li}_2(y+z)+\pi^2-25\right].
\end{eqnarray}

With the same rules as those stated in the previous subsections, the
differences between other $\gamma_5$ strategies/regularization
schemes and the FDH scheme in KKS $\gamma_5$ scheme are given by
\begin{eqnarray}
\left(\delta|\overline{\mathcal{M}}_v|^2+\delta|\overline{\mathcal{M}}_{ct}|^2\right)^{KKS}_{HV/CDR}&=&|\overline{\mathcal{M}}_{0}|^2
\frac{\alpha_s~C_F}{4\pi\Gamma(1-\epsilon)}\left(\frac{4\pi\mu^2}{m_t^2}\right)^{\epsilon}\nonumber\\&&
\left[\frac{3(1-y-z)(y+z)}{y(1-y)}\frac{1}{\epsilon}-\frac{1}{2y(1-y)}\right.\nonumber\\&&
\left(5y^2+22yz+11z^2-5y-11z\right.\nonumber\\&&\left.\left.+4(1-y-z)(y+z)\ln[(1-y-z)(y+z)]\right)\right],\nonumber\\
\left(\delta|\overline{\mathcal{M}}_v|^2+\delta|\overline{\mathcal{M}}_{ct}|^2\right)^{BMHV}_{FDH}&=&|\overline{\mathcal{M}}_{0}|^2
~\frac{\alpha_s~C_F}{\pi},\nonumber\\
\left(\delta|\overline{\mathcal{M}}_v|^2+\delta|\overline{\mathcal{M}}_{ct}|^2\right)^{BMHV}_{HV/CDR}&=&|\overline{\mathcal{M}}_{0}|^2
~\frac{5~\alpha_s~C_F}{4~\pi}.
\end{eqnarray}

To check our results, we also treat the numerators of loop
amplitudes in four-dimensions by adding the $R_2$ terms at last. All
of the results discussed above are recovered using this method.
Because of the right-handed currents\cite{Shao:2011tg} of the $R_2$
in the BMHV $\gamma_5$ scheme, the unrenormalized virtual
contributions are the same within the same $\gamma_5$ treatment, and
only the renormalization constants are different.

The remaining regularization scheme-dependent terms should be
canceled by the real radiation part. The scheme-dependent terms in
real corrections resulted from the soft and collinear region of
phase space. The two cutoff phase space slicing method given by B.Harris and J.Owens is used
here\cite{Harris:2001sx}. The analytical result within the FDH and
KKS regularization scheme is given by
\begin{eqnarray}
\left(|\overline{\mathcal{M}}_{sc}|^2\right)^{KKS}_{FDH}&=&|\overline{\mathcal{M}}_{0}|^2
\frac{\alpha_s~C_F}{4\pi\Gamma(1-\epsilon)}\left(\frac{4\pi\mu^2}{m_t^2}\right)^{\epsilon}\nonumber\\&&
\left[\frac{6}{\epsilon^2}-\frac{4\ln[(1-y-z)(y+z)]-11}{\epsilon}
+2\ln^2[\frac{y~z}{1-y-z}]\right.\nonumber\\&&-2\ln^2(1-y)-2\ln^2(1-z)-2\ln^2(y+z)\nonumber\\&&
+4\ln[(1-y)(1-z)(y+z)]\ln(\frac{\delta_s}{\delta_c})-9\ln(\delta_c)\nonumber\\&&-4\ln(\delta_s)
-12\ln(\delta_s)\ln(\delta_c)+8\ln[(1-y-z)(y+z)]\ln(\delta_s)\nonumber\\&&
\left.+6\ln^2(\delta_s)+4~\text{Li}_2[-\frac{1-y-z}{y~z}]+22-\frac{7\pi^2}{3}\right],
\end{eqnarray}
where $\delta_s$ and $\delta_c$ are two parameters to isolate the
soft and collinear regions, respectively. The differences between
other regularization schemes and the above scheme are
\begin{eqnarray}
\left(\delta|\overline{\mathcal{M}}_{sc}|^2\right)^{KKS}_{HV/CDR}&=&-|\overline{\mathcal{M}}_{0}|^2
\frac{\alpha_s~C_F}{4\pi\Gamma(1-\epsilon)}\left(\frac{4\pi\mu^2}{m_t^2}\right)^{\epsilon}\nonumber\\&&
\left[\frac{3(1-y-z)(y+z)}{y(1-y)}\frac{1}{\epsilon}-\frac{1}{2y(1-y)}\right.\nonumber\\&&
\left(5y^2+22yz+11z^2-5y-11z\right.\nonumber\\&&\left.\left.+4(1-y-z)(y+z)\ln[(1-y-z)(y+z)]\right)\right]\nonumber\\
&=&-\left(\delta|\overline{\mathcal{M}}_v|^2+\delta|\overline{\mathcal{M}}_{ct}|^2\right)^{KKS}_{HV/CDR},\nonumber\\
\left(\delta|\overline{\mathcal{M}}_{sc}|^2\right)^{BMHV}_{FDH}&=&0,\nonumber\\
\left(\delta|\overline{\mathcal{M}}_{sc}|^2\right)^{BMHV}_{HV/CDR}&=&|\overline{\mathcal{M}}_{0}|^2
~\frac{3~\alpha_s~C_F}{4~\pi}.
\end{eqnarray}

In the BMHV $\gamma_5$ scheme, we also obtain the scheme-independent
results after including the finite renormalization to the
axial-vector currents. This was done in order to maintain the Ward
identities as already shown in the last two subsections.

In the hard noncollinear phase space region, we treat the squared
matrix element of $t(p_t)\rightarrow b(p_b)u(p_u)d(p_d)g(p_g)$ in
four dimensions. Dimensionless variables are redefined as follows:
\begin{eqnarray}
(p_t-p_g)^2&=&m_t^2~x,(p_t-p_u)^2=m_t^2~y,(p_t-p_d)^2=m_t^2~z,\nonumber\\
(p_u+p_d)^2&=&m_t^2~k,(p_u+p_g)^2=m_t^2~l,r=\frac{m_W}{m_t},w=\frac{\Gamma_W}{m_W}.
\end{eqnarray}
The averaged squared amplitude is
\begin{eqnarray}
\label{eq1}
|\overline{\mathcal{M}}_{real}|^2&=&\frac{3~e^4~g_s^2~C_F}{|s_w|^4~m_t^2}\left\{\frac{1}{(k-r^2)^2+(wr^2)^2}
\frac{1}{(1-x)^2(1-y-z-k)}\right.\nonumber\\&&\left[\left(x-3\right)\left(k+l\right)^2k+
2\left(x z-2x-y-4z+4\right)k^2\right.\nonumber\\&&+\left(\left(2 x
z-5x-4y-10z+11\right)l-x^2-y^2-7z^2+4y-4y
z+15z\right.\nonumber\\&&\left.+\left(y^2-2y+z^2-7z+6\right)x-7\right)k+
\left(1-z\right)\left(2-x-y-z\right)\left(1-x-2z\right)\nonumber\\&&
\left.-\left(x+2 y+2 z-3\right)l^2-\left(x+2 z-2\right)\left(x+2 y+2
z-3\right)l\right]\nonumber\\&&+\frac{1}{(2-x-y-z-r^2)^2+(wr^2)^2}\frac{1}{l
(x+y+z+k+l-2)}\nonumber\\&&\left[\left(2-x-y-z\right)\left(1-z-k-l\right)^2-\left(k+l\right)
\left(1-z-k-l\right)\right.\nonumber\\&&
+\left.\left.\left(1-y\right)\left(l-y\left(2-x-y-z\right)\right)\right]\right\}
\end{eqnarray}
There are two kinds of Breit-Wigner distributions of the W boson in
Eq.(\ref{eq1}). The first term originated from the first two real
diagrams, while the second is contributed by the last two final
state radiative diagrams. Because of color flow, there is no
interference observed between the first two and the last two
diagrams.
\section{Jet Algorithms And Phase Space}
At high energy colliders, it was pointed out that the observed jets
provided a view of parton (e.g. gluon and quark) interactions
occurring at short distances\cite{Sterman:1977wj}. At leading-order
(LO) level, partons can be naively treated as jets, while at NLO
level this coarse treatment often suffers from soft and collinear
divergences. Therefore, an infrared-collinear safe jet definition is
necessary in investigating strong interaction physics. Nowadays,
these jet definitions play important roles in collider physics.
Following the jet definition description in
Refs.\cite{Catani:1993hr,Catani:1992zp}, the requirements
implemented in a jet algorithms are as follows:
\begin{itemize}
\item simple to use in experiments and theoretical calculations,
\item infrared and collinear safe,
\item small hadronization corrections.
\end{itemize}
At hadron colliders, a well-defined jet algorithm must be able to
factorize initial-state collinear singularities; they should also be
isolated from the contamination of hadron remnants and underlying
soft events.

Since the advent of jet production in electron-positron and hadron
colliders, it has become one main tool in QCD research. Many kinds
of algorithms have been proposed and developed. Essentially, the two
classes of jet algorithms present mainly the clustering
algorithms\cite{Bartel:1986ua,Bethke:1988zc} and the cone-type
algorithms\cite{Sterman:1977wj,Abe:1991ui,Arnison:1983gw,Salam:2007xv}.
In the present study, we focused on the three popular inclusive
clustering algorithms, namely the $k_{\perp}$-clustering
algorithm\cite{Catani:1993hr,Catani:1992zp,Ellis:1993tq},the $\rm
Cambridge/Aachen$ clustering
algorithm(CA)\cite{Dokshitzer:1997in,Wobisch:1998wt} and the
anti-$k_{\perp}$ clustering algorithm\cite{Cacciari:2008gp}
respectively. These three inclusive clustering algorithms can be
described uniformly:
\begin{itemize}
\item Define a distance $d_{ij}=min(p_{Ti}^{2r},p_{Tj}^{2r})\frac{\Delta
R_{ij}}{R}$ between each pair of protojets i and j, as well as a
distance $d_{iB}=p_{Ti}^{2r}$ between each protojet i and the
beam,with $r=+1,0,-1$ corresponding to $k_{\perp}$,CA, and
anti-$k_{\perp}$ respectively.

\item Find the smallest of all the $d_{ij}$ and $d_{iB}$ and
label it as $d_{min}$.

\item If $d_{min}$ is a $d_{ij}$, then cluster protojets i and j as a new
protojet with a selected combination procedure. If the distance
between protojet i and the beam is the shortest, set the protojet i
aside and leave it without any further clustering as a possible jet
candidate.

\item Repeat the items above until there is no protojet left.

\item Perform some cuts (as in the experiment) to select
jet(s) of interest.
\end{itemize}
Here $\Delta
R_{ij}=\sqrt{\left(\eta_{i}-\eta_{j}\right)^2+\left(\phi_{i}-\phi_{j}\right)^2}$
($\eta$ and $\phi$ are rapidity and azimuthal angle respectively).
As $E$ can be measured at $e^+e^-$ colliders rather than only $p_T$
at hadronic colliders, one should use $E$ instead of $p_T$ and
$\Delta S_{ij}=\sqrt{\Delta
\theta_{ij}^2+\left(\sin{\frac{\theta_i+\theta_j}{2}}\Delta
\phi_{ij}\right)^2}$ instead of $\Delta R_{ij}$ at $e^+e^-$
colliders.

It was also emphasized in Ref.\cite{Ellis:1993tq} that traditional
cone-type jet algorithms were related to clustering algorithms by
the approximation $R_{cluster}=1.35\times R_{cone}$.

In the present study, we only used the three clustering algorithms
with the E-scheme recombination to reconstruct our leading two jets
from top quark hadronic decay in the next section (one can also use
other recombination procedures as suggested in
Ref.\cite{Catani:1993hr} and references therein). In addition, we
used hadron collider clustering algorithms and electron-positron
collider clustering algorithms but without any cut in our
calculation.

The last topic of this section is about a phase-space integration
treatment. Given that we should reconstruct the four momenta of all
final states in order to reconstruct two leading jets, we built up
the n-particle phase space iteratively by nested integration over
invariant masses and solid angles of outgoing particles, similar to
the strategy in Ref.\cite{Hahn:2006qw}.
\section{Results}
The dijet invariant mass distributions with different clustering jet
algorithms are presented in this section.

As discussed in the previous section, we used two variations of
clustering jet algorithms in our top decay process in the c.m. frame
of the top quark. In these two variations, we chose the distances
defined at hadron colliders ( i.e. use $p_T$ ) and $e^+e^-$
colliders( i.e. use $E$), respectively, to reconstruct the final
jets. Afterward, two leading jets in energy $E$ were chosen to
construct their invariant mass $m_{jj}$. Here,we call the first
types KT1, CA1, anti-KT1, while the second types are denoted as KT2,
CA2, anti-KT2.

The following input parameters are used:
\begin{eqnarray}
\alpha^{-1}&=&129, \alpha_s(m_Z)=0.119,\nonumber\\
m_W&=&80.399~GeV, \Gamma_W=2.085~GeV, m_Z=91.1876~GeV,
\Gamma_Z=2.4952~GeV,\nonumber\\
s_w^2&=&1-\frac{m_W^2-i~m_W\Gamma_W}{m_Z^2-i~m_Z\Gamma_Z}=0.222657-1.11098\times10^{-3}i,
\end{eqnarray}
with two groups of top quark mass and renormalization scale $\mu$
choices, i.e. $m_t=175~GeV$, $\mu=80.4~GeV$ and $m_t=172.5~GeV$,
$\mu=m_t=172.5~GeV$\footnote{As shown in Sec.\uppercase\expandafter{\romannumeral 3}, the only scale $\mu$
dependence in $\mathcal{O}(\alpha^2\alpha_s)$ is $\alpha_s(\mu)$,
which is just a global factor and does not change our dijet
invariant mass distribution significantly. However, the top quark
mass dependence in our results is much more complicated. Therefore,
we choose two top quark mass benchmark points to investigate its
influence on our curves' shape, and do not plot the scale dependence
in this paper.}.

\begin{center}
\begin{figure}
\hspace{0cm}\includegraphics[width=14cm]{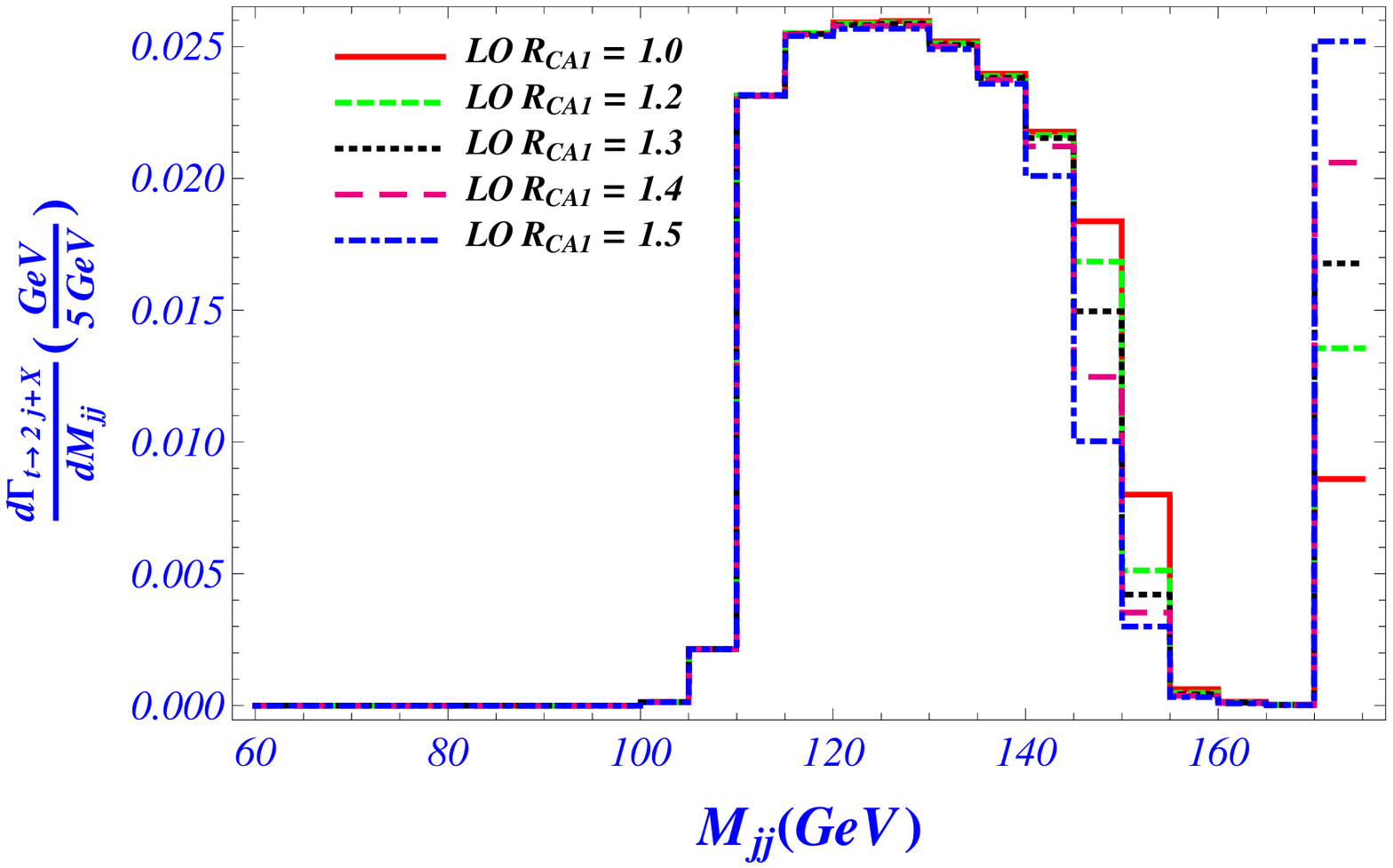}
\hspace{0cm}\includegraphics[width=14cm]{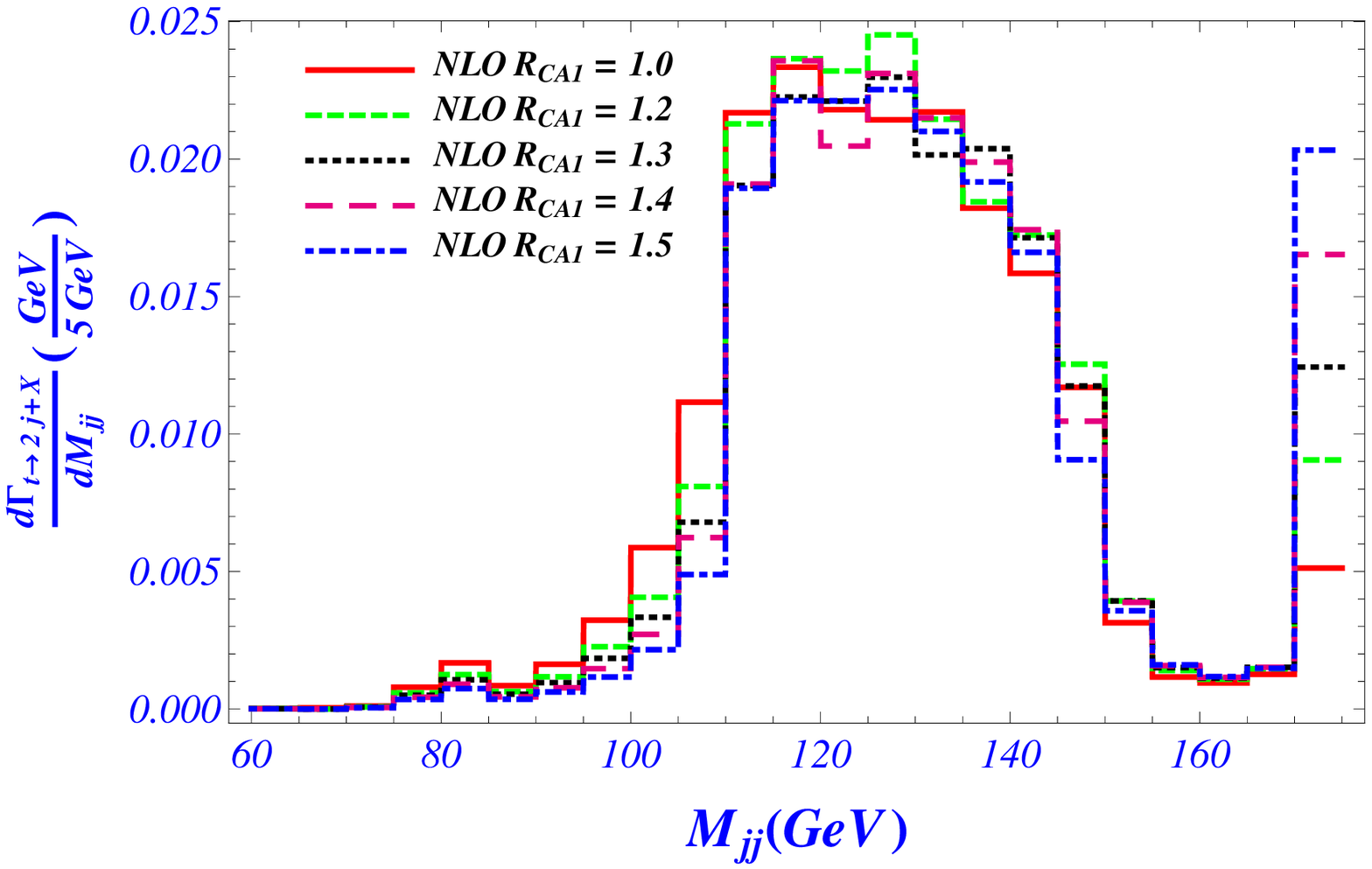}
\caption{\label{fig:LO_R} The LO (upper panel) and NLO (lower panel)
dijet invariant mass distribution from top decay with different $R$
using the ${\rm CA1}$ clustering jet algorithm ($m_t=175~GeV,
\mu=80.4~GeV$). Plotted are R=1.0 (solid line), 1.2 (short-dashed
line), 1.3 (dotted line), 1.4 (long-dashed line), and 1.5
(dot-dashed line), respectively.}
\end{figure}
\end{center}

We varied the parameter $R$ in the ${\rm CA1}$ jet algorithm and
compared its influence on our results in Fig.\ref{fig:LO_R}. Only
when $R\geq1.0$ in this algorithm, the infrared and collinear safety
in each bin is maintained. Therefore, to ensure reliability, we
choose $R=1.4$ in the first type jet algorithms and $R=1.3$ in the
second ones for the rest of the paper. There are some interesting
characters in these two figures. The variation of $R$ slightly
changed our domain region (110-150 GeV) both in LO and NLO level.
The larger $R$ reconstructs a smaller number of final jets; it makes the
number of events in the last bin (170-175~GeV) larger with larger
$R$. At LO, the distributions dropped sharply below 110 GeV and
vanish below 100 GeV, as shown on the upper panel of
Fig.\ref{fig:LO_R}. In contrast, a NLO QCD correction resulted in the
smooth descent of the low energy tail. The peak in
Fig.\ref{fig:LO_R} (lower panel) between 80 GeV to 85 GeV is the W
boson's resonance.

\begin{center}
\begin{figure}
\hspace{0cm}\includegraphics[width=14cm]{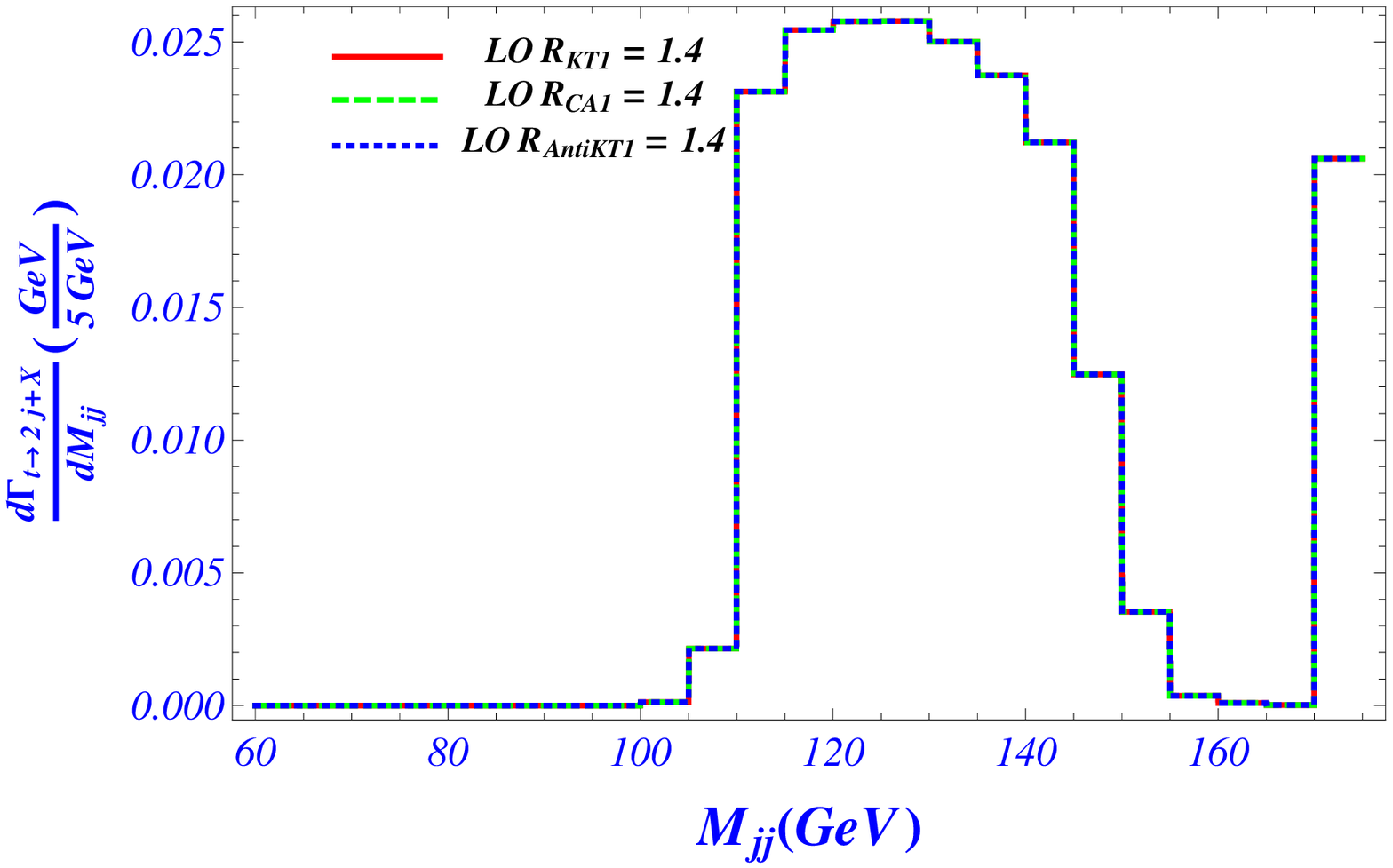}
\hspace{0cm}\includegraphics[width=14cm]{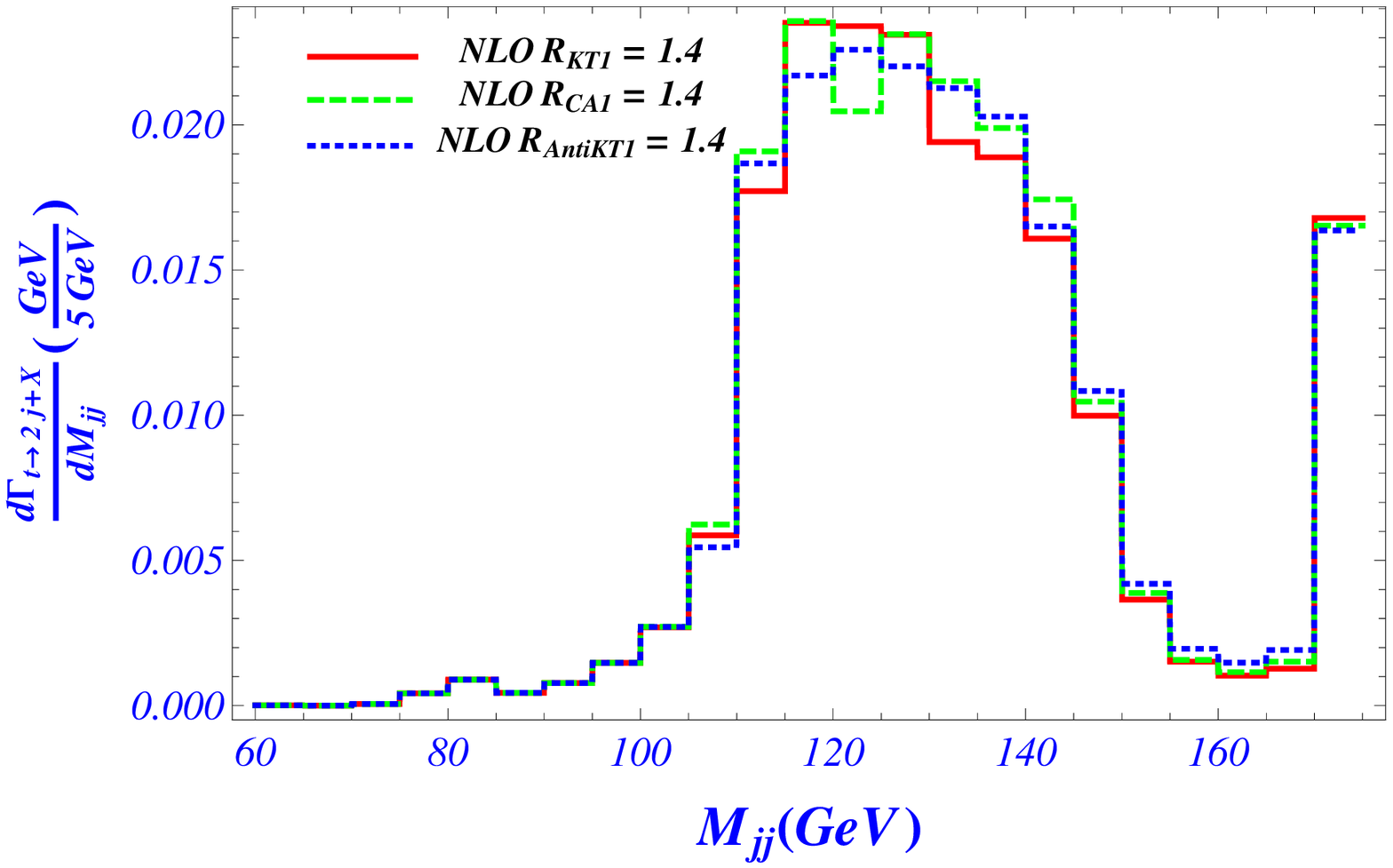}
\caption{\label{fig:Clus_1} The influence on distribution with
different clustering jet algorithms of the first type ($m_t=175~GeV,
\mu=80.4~GeV$). LO is in the upper panel, while the lower panel is
for the NLO results. Plots are KT1 (solid line), CA1 (dashed line),
and anti-KT1 (dotted line).}
\end{figure}
\end{center}

\begin{center}
\begin{figure}
\hspace{0cm}\includegraphics[width=14cm]{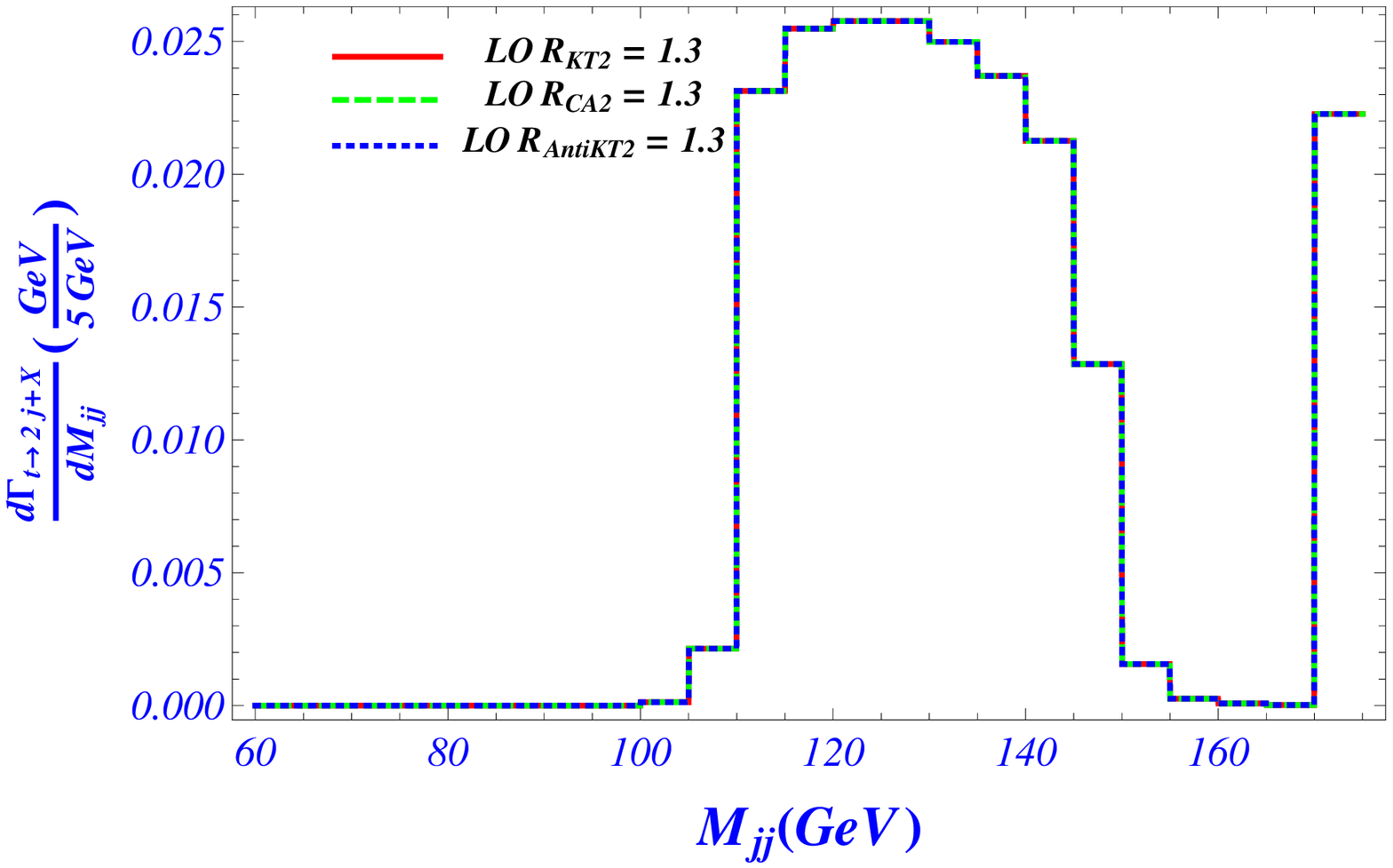}
\hspace{0cm}\includegraphics[width=14cm]{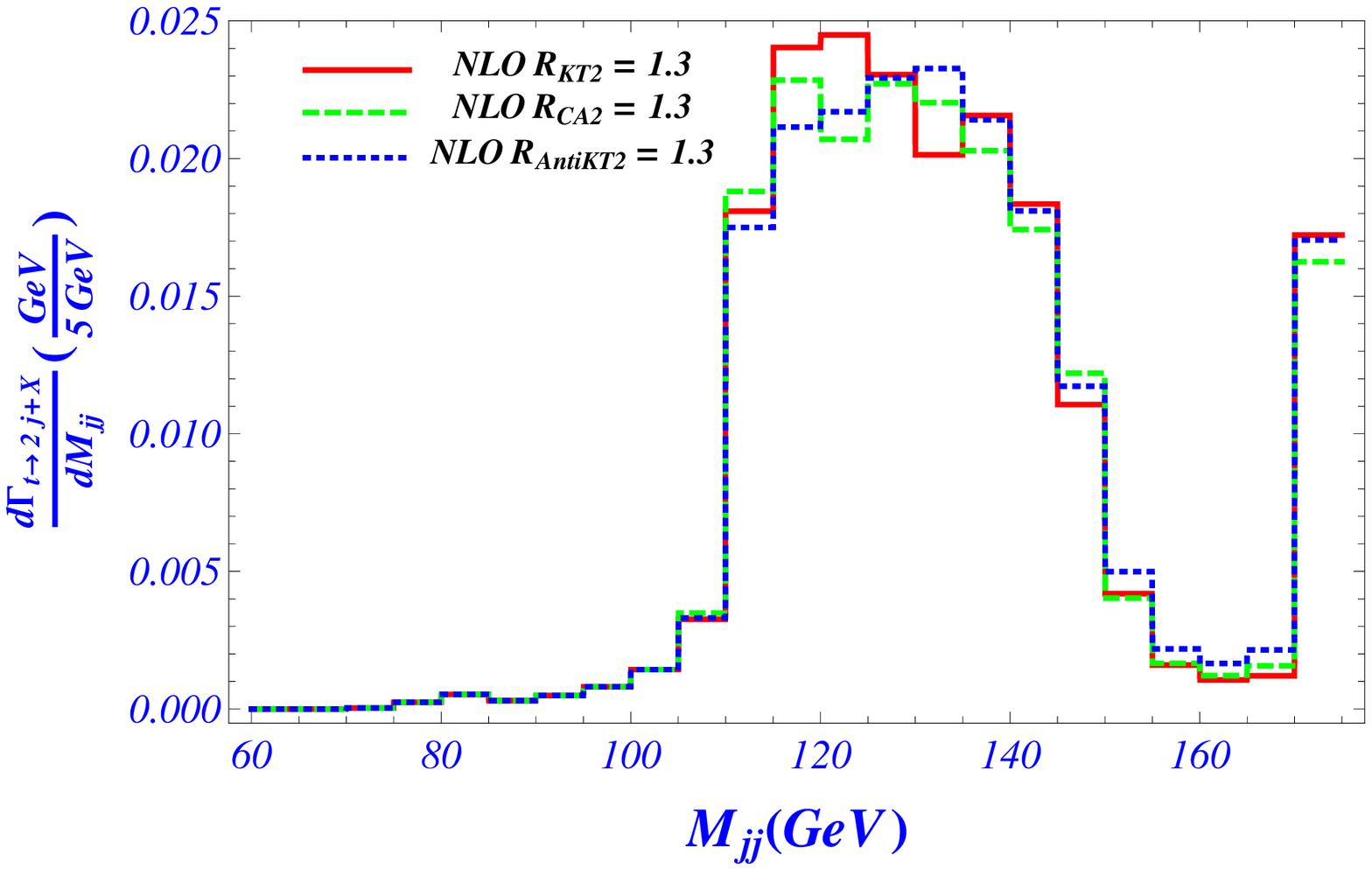}
\caption{\label{fig:Clus_2} The influence on distribution with
different clustering jet algorithms of the second type
($m_t=175~GeV,\mu=80.4~GeV$). LO is in the upper panel, while the
lower panel is for the NLO results. Plots are KT2 (solid line), CA2
(dashed line), and anti-KT2 (dotted line).}
\end{figure}
\end{center}

\begin{center}
\begin{figure}
\hspace{0cm}\includegraphics[width=14cm]{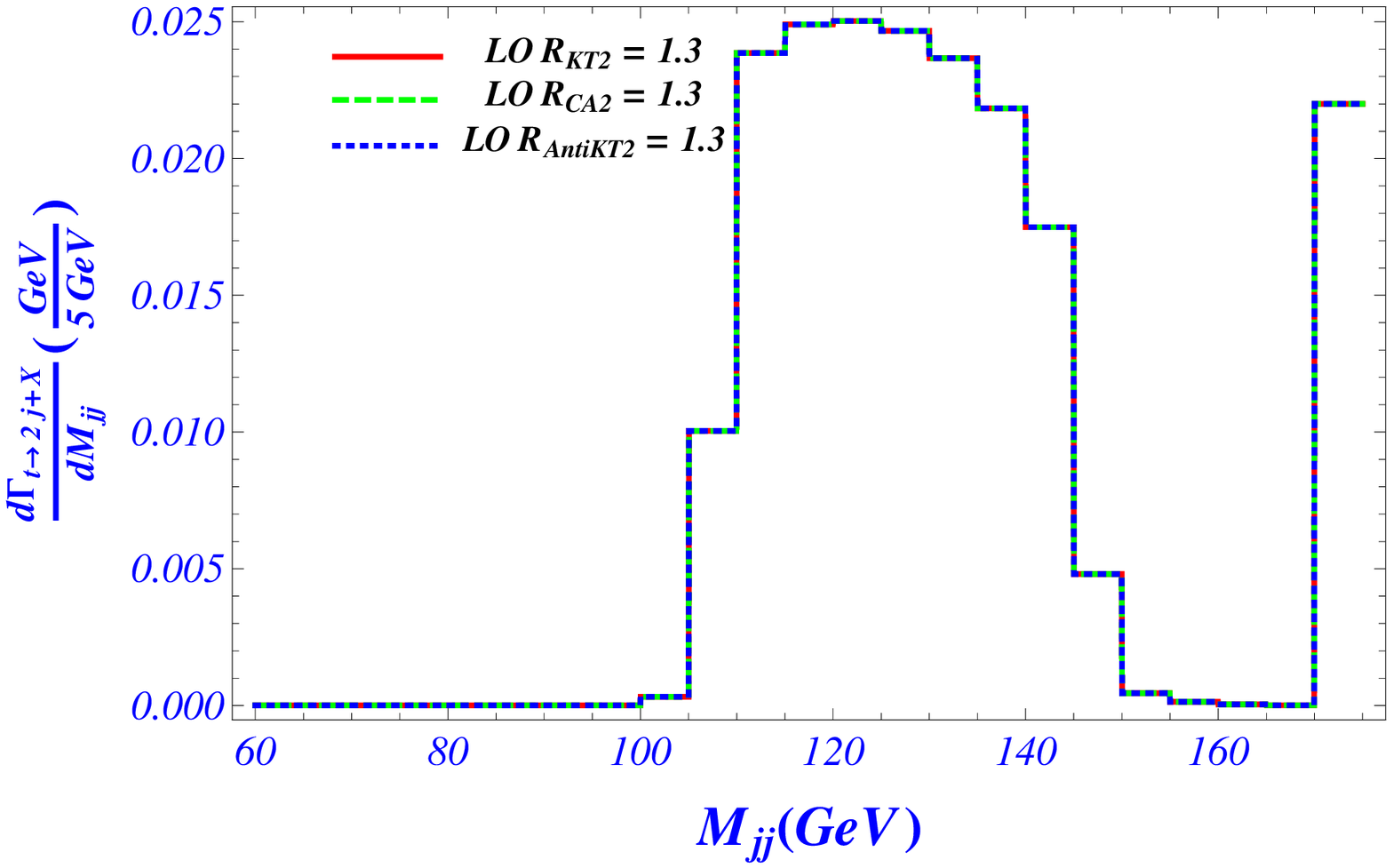}
\hspace{0cm}\includegraphics[width=14cm]{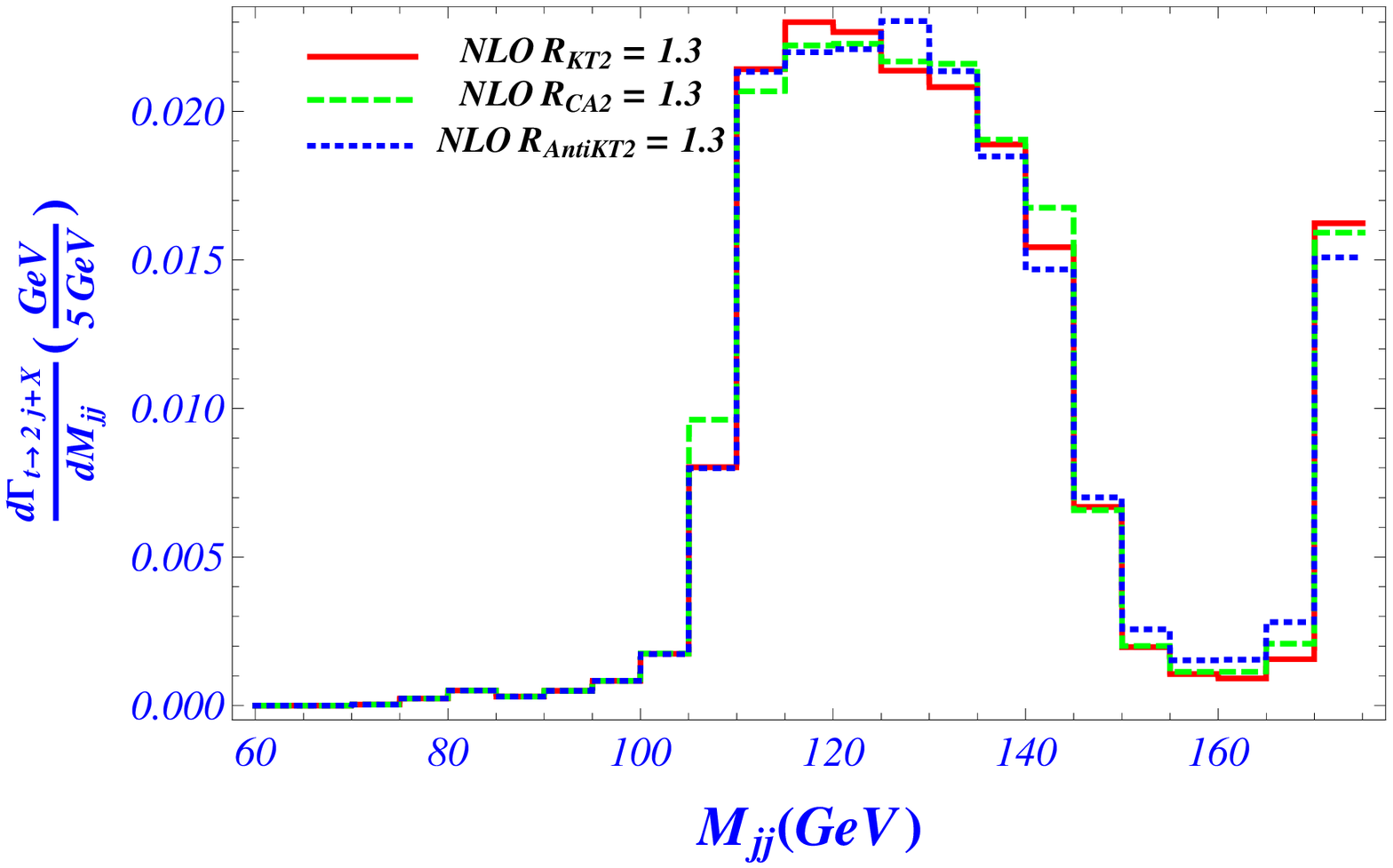}
\caption{\label{fig:Clus_3} The influence on distribution with
different clustering jet algorithms of the second type
($m_t=172.5~GeV, \mu=172.5~GeV$). LO is in the upper panel, while
the lower panel is for the NLO results. Plots are similar to
Fig.\ref{fig:Clus_2}.}
\end{figure}
\end{center}

Histograms in Figs.\ref{fig:Clus_1} and \ref{fig:Clus_2} establish
the influences of clustering jet algorithms to the dijet invariant
mass distribution. The LO distributions reconstructed by various
algorithms are almost indistinguishable. In comparison,there are
some differences in the substructures of NLO histograms. The
combination sequence of protojets is responsible for these tiny
distinctions\footnote{Statistical uncertainties are also responsible
for these differences in the histograms. They change our results by
about 4 percent.}. Soft protojets may be clustered before the
hard ones in $k_{\perp}$, while the situation may be totally
different in anti-$k_{\perp}$. For comparison,we also plot the
histograms with $m_t=172.5 GeV$ and $\mu=172.5 GeV$ in
Fig.\ref{fig:Clus_3}.

\section{Conclusions}

We have performed  QCD radiative corrections to the dijet production
in the unpolarized top quark hadronic decay in the complex mass
scheme. We carefully checked the independence of dimensional
regularization schemes and $\gamma_5$ strategies in our analytical
formalism. Applying different clustering jet definitions, we
obtained our final dijet invariant mass distributions. The obtained
dijet mass distributions from the top quark decay are useful to
understand the top quark properties and also to distinguish these
dijets from those produced via other sources. Therefore, these
results are useful in investigating the recent CDF $Wjj$ anomaly and
clarifying this interesting issue.  Furthermore, a more careful
investigation for top and W boson associated production at hadron
colliders will be definitely needed.

\begin{acknowledgments}
We are grateful to K. Wang for the help in some program techniques.
We also thank J. Gao, C. Meng, and Y.Q. Ma for useful discussions.
This work was supported by the National Natural Science Foundation
of China (No.10805002, No.11021092, No.11075002, No.11075011), the
Foundation for the Author of National Excellent Doctoral
Dissertation of China (Grant No. 201020), and the Ministry of
Science and Technology of China (2009CB825200).

\end{acknowledgments}


\end{document}